\documentstyle[aps,prl,epsf,twocolumn,epsfig,floats]{revtex}

\begin{document}
\draft



\wideabs{

\title{Electrodynamics of a Coulomb Glass in n-type Silicon}
\author{E. Helgren, N. P. Armitage and G. Gr\"{u}ner}
\address{Dept. of Physics and Astronomy, University of California Los Angeles, Los Angeles, CA 90095}

\date{\today}
\maketitle
\begin{abstract}

Optical measurements of the real and imaginary frequency dependent
conductivity of uncompensated n-type silicon are reported. The
experiments are done in the quantum limit, $ \hbar\omega >
k_{B}T$, across a broad doping range on the insulating side of the
Metal-Insulator transition (MIT). The observed low energy linear
frequency dependence shows characteristics consistent with
theories of a Coulomb glass, but discrepancies exist in the
relative magnitudes of the real and imaginary components. At
higher energies we observe a crossover to a quadratic frequency
dependence that is sharper than expected over the entire dopant
range. The concentration dependence gives evidence that the
Coulomb interaction energy is the relevant energy scale that
determines this crossover.

\end{abstract}
} 

Doped semiconductors close to the MIT are systems ideally suited
for the study of strong electron-electron interactions. Below a
critical doping the material is an insulator, i.e. at zero
temperature there is no DC conductivity. The charge carriers are
localized in the Anderson sense, and screening is reduced from the
metallic regime.

Anderson coined the term Fermi glass to describe a non-interacting
disordered insulating system whose universal properties,
independent of system specific details, are determined by Fermi
statistics alone \cite{Anderson}. Mott's original treatment of
such a system \cite{Mott}, which has a finite density of states at
the Fermi level, but where localization makes an insulator, did
not address the issue of electron-electron interactions as pointed
out by Pollak \cite{Pollak}, and then Efros and Sklovskii (ES)
\cite{ES85}. A Fermi glass that includes interactions between
localized electrons has been termed a Coulomb glass. Among other
things, such a system is typified by a depletion in the single
particle density of states around the Fermi level which was termed
by ES the Coulomb gap, $\Delta$ \cite{ES75}.

By taking into account the mean Coulomb interaction between two
sites forming a resonant pair $U(r_{\omega}) =
e^{2}/\varepsilon_{1} r_{\omega}$, where $r_{\omega} =
\xi[ln(2I_{0}/\hbar\omega)]$ is the most probable hop distance
between pairs, $I_{0}$ is the hopping attempt rate and
$\varepsilon_{1}$ is the dielectric constant, ES derived the real
part of the ac conductivity of a Coulomb glass to be:
\begin{equation}
    \sigma_{1} = \beta e^{2}g_{0}^{2}\xi^{5}\omega[ln(2I_{0}/\hbar\omega)]^{4}[\hbar\omega +
    U(r_{w})].
    \label{eq:ESxover}
\end{equation}
Here $\beta$ is a constant of order one, $g_{0}$ is the
non-interacting density of states and $\xi$ is the localization
length. This formula takes on a different frequency dependence in
two limits. When the photon energy, $\hbar\omega
> U(r_{\omega})$, one recovers the same quadratic frequency
dependence that Mott derived for a non-interacting Fermi glass.
Here the Coulomb glass is indistinguishable from the Fermi glass
in so far as the high frequency limit of the conductivity is
concerned. In the opposite limit, $\hbar\omega < U(r_{\omega})$,
the conductivity of a Coulomb glass will show an approximately
linear dependence on frequency, plus logarithmic corrections. The
imaginary component of the complex conductivity, $\sigma_{2}$, as
predicted by Efros\cite{Efros85}, should be identical to
$\sigma_{1}$ up to a logarithmic factor. We should note that Eq.
(\ref{eq:ESxover}) was derived for the case where $\hbar\omega >
\Delta$, the Coulomb gap width. However a linear dependence
(albeit with additional logarithmic corrections) and an eventual
crossover to Mott's non-interacting quadratic law is still
expected even for the case where $\hbar\omega < \Delta$.

Despite its fundamental importance, there have been very few AC
conductivity studies that have been done in the quantum limit $
\hbar\omega > k_{B}T$, yet at low enough frequencies to probe the
relevant energy ranges of Eq. (\ref{eq:ESxover}). Only very
recently, have measurements been attempted that address these
issues. M. Lee et al. found that for concentrations close to the
MIT the expected linear to quadratic crossover occurs, but is much
sharper than predicted \cite{MLee01}. Other recent work was done
on amorphous NbSi, where the general systematics of a linear in
frequency Coulomb glass response was found, but a crossover to the
non-interacting quadratic dependence was not reached
\cite{Helgren}.

In this Letter we report the first measurements of the frequency
dependent real and imaginary conductivity on the insulating side
of the MIT for a doped, crystalline semiconductor. We find a
concentration and frequency dependence of the components
consistent with predictions of a Coulomb glass, but a discrepancy
arises in the predicted ratio of their relative magnitudes. At
higher frequencies a crossover to quadratic Mott-like behavior is
observed and due to having measured across a broad dopant range,
the concentration dependence of the crossover energy is shown to
be consistent with the Coulomb interaction energy $U$ and not the
Coulomb gap width $\Delta$.

Nominally uncompensated n-type silicon samples were obtained from
Recticon Enterprises Inc. A boule of silicon was grown using the
Czochralski method and sliced into 1 mm thick slabs. Room
temperature resistivity was measured using an ADE 6035 resistivity
gauge, and the concentration determined using the Thurber
scale\cite{Thurber}. A number of samples were etched with a $4\%
HF + 96\% HNO_{3}$ solution; this resulted in no difference in the
results of frequency dependent conductivity measurements. The Si:P
samples discussed in this Letter span a range from 39\% to 69\%,
stated as $\frac{x}{x_{c}} $, a percentage ratio of the sample's
dopant concentration to the critical concentration at the MIT.

\begin{figure}[tbh]
\centerline{\epsfig{figure=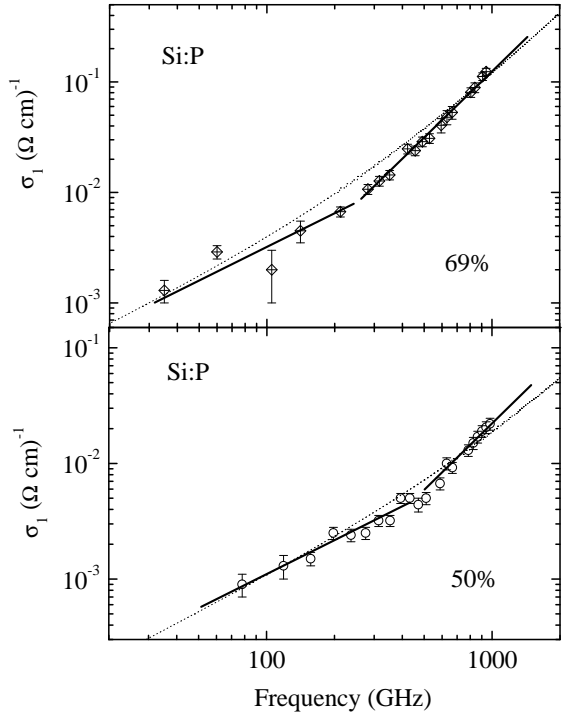,width=8cm}} \vspace{.2cm} \caption{T = 2.8 K real part of the frequency
dependent conductivity versus frequency. The Si:P samples shown are at 50\% and 69\% dopant concentration relative
the critical concentration, $x_{c} = 3.5 \times 10^{18} cm^{-3}$. The solid lines are linear and quadratic fits to
the lower and upper portions of the data respectively. The dotted line is a fit to the form of Eq. (1).}
\end{figure}

At 35 and 60 GHz the real part of the conductivity was evaluated
from the measured loss of highly sensitive resonant cavities via
the perturbation method. The technique and analysis is well
established\cite{Gruner}. In the millimeter spectral range, 80 GHz
to 1 THz, backward wave oscillators were employed as coherent
sources in a transmission configuration\cite {Schwartz}.
Fabry-Perot like resonances in the transmission were analyzed
uniquely determining both components of the complex conductivity.
Cavity measurements were performed down to 1.2 K and in the
millimeter spectral range down to 2.8 K. With these base
temperatures the quantum limit $ \hbar\omega > k_{B}T$ ($1 K
\approx 20 GHz$) of the system was being investigated.

\begin{figure}[tbh]
\centerline{\epsfig{figure=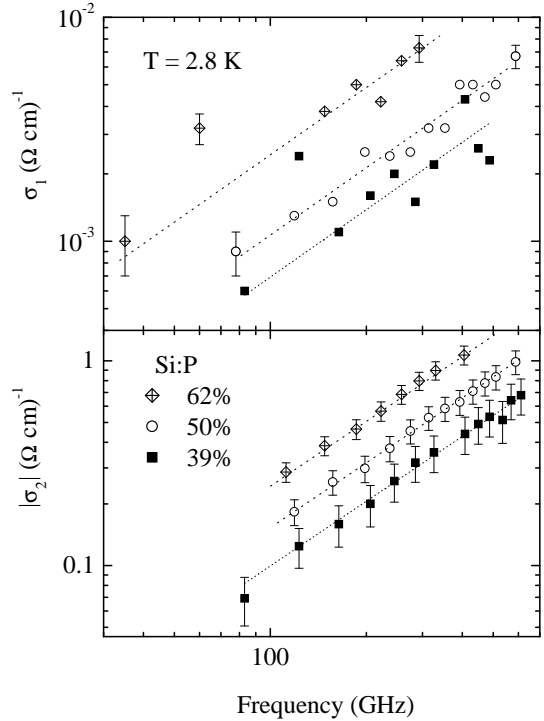,width=8cm}} \vspace{.2cm} \caption{The real part and the magnitude of the
imaginary part of the optical conductivity for three different Si:P samples on the insulating side of the MIT is
shown. The stated percentages are the sample's dopant concentration relative the critical concentration. The dotted
lines are linear best fits showing the trend of increasing conductivity as the MIT is approached. The error bars
shown for $\sigma_{1}$ are representative of those for all the data points. Some of the lower frequency points were
determined using resonant cavities at a base temperature of 1.2 K. }
\end{figure}

The real part of the frequency dependent conductivity,
$\sigma_{1}$, at $T=2.8\;K$ for two samples is shown in Fig. 1.
This data, representative of all the samples, shows an
approximately linear dependence at low frequencies and then a
sharp crossover to an approximately quadratic behavior at higher
frequencies. This is the qualitatively expected behavior from Eq.
(\ref{eq:ESxover}). However, as seen by the overlayed fits, Eq.
(\ref{eq:ESxover}) provides only a rough guide. The solid lines
are linear and quadratic fits to the low frequency and high
frequency data respectively. Individually these functions fit the
data well, however the full crossover function does not. The
dotted line is a fit using the same method as Ref. \cite{MLee01},
namely forcing the linear portion to pass through the origin as
well as the low frequency data and leaving the pre-factor of the
quadratic term as a free variable. The crossover between linear
and quadratic portions is much more abrupt than the ES function
predicts and is observed over our entire doping range as was
observed previously in an analogous system, Si:B, for samples
closer to the MIT \cite{MLee01}.

Having shown that the frequency dependence is qualitatively
consistent with a Coulomb glass, we focus on the low frequency
regime. Both the real part and the magnitude of the imaginary
part, $|\sigma_{2}|$. of the complex conductivity are plotted as a
function of frequency in Fig. 2. The dotted lines are linear fits
and show the relative increase in the magnitude of both components
as the MIT is approached.

A complex conductivity obeying a power law can be expressed in a
simple Kramers-Kronig compatible form, $\sigma_{1}(\omega) + i\;
\sigma_{2} (\omega) = A (i \omega)^{\alpha}$, in order to have an
independent means for determining the exponent. To determine the
power $\alpha$ one can utilize the ratio of $|\sigma_{2}|$ versus
$\sigma_{1}$ (with the frequency as a variable). The power
$\alpha$ is given by,

\begin{equation}
    \ \alpha = \frac{2}{\pi} tan^{-1} \left( \frac{|\sigma_{2}|}{\sigma_{1}} \right).
    \label{eq:alpha}
\end{equation}

\begin{figure}[tbh]
\centerline{\epsfig{figure=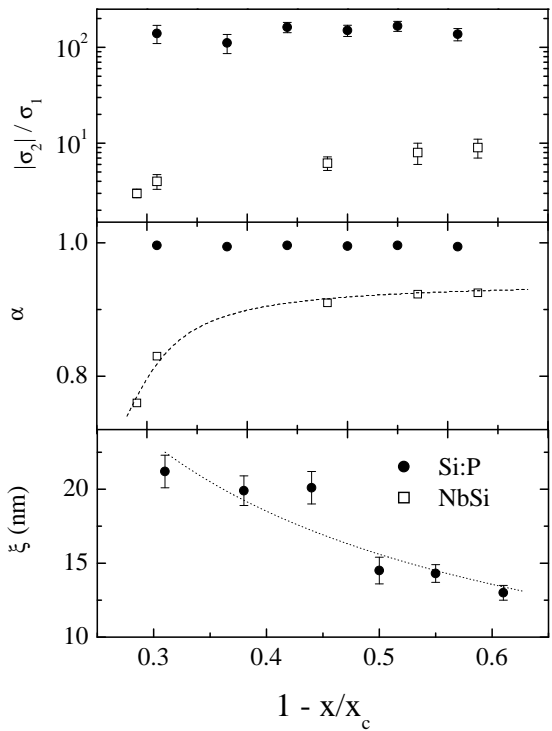,width=8cm}} \vspace{.2cm} \caption{The upper panel shows the ratios of the
magnitude of the imaginary to the real part of the complex conductivity for samples of Si:P and amorphous NbSi. The
NbSi data is adapted from Ref. 8. The middle panel shows the calculated powers of $\alpha$ as determined from Eq.
6. The dashed line through the NbSi data is a guide to the eye.  The bottom panel shows the divergence of the
localization length (using $I_{0} = 10^{13} s^{-1}$), and the dotted line is a power law fit of
$(1-x/x_{c})^{-\beta}$, where $\beta$ is found to be 0.95.}
\end{figure}

The uppermost panel in Figure 3 shows the ratio of the imaginary
to the real part of the conductivity for Si:P. NbSi is an
amorphous insulating glass which is characterized by a vastly
higher density of states (DOS) at $E_{F}$ \cite{Helgren}. Disorder
should give a somewhat similar functional dependence for the AC
conductivity, but with very different energy scales. Data from it
is included as another example of AC conductivity in a disordered
system and for comparison purposes, as Si:P has a very small DOS
in its impurity band. First we note that this ratio for Si:P
remains large and approximately constant across our range of
dopant concentrations. From theory, one expects $|\sigma_{1}|$ to
be approximately equal to $|\sigma_{2}|$ to within a factor of 2-5
(with a reasonable estimate for $I_{0}$) as predicted by Efros
\cite{Efros85}. Applied to Si:P, Eq. (\ref{eq:ESxover}) in the
$\hbar\omega < U(r_{\omega})$ limit correctly predicts a linear
correspondence between $\sigma_{1}$ and $|\sigma_{2}|$, but the
theory incorrectly predicts the measured proportionality by at
least a factor of thirty. The proportionality is closer for NbSi
but has a dependence on the doping concentration. This may be
related to entering the quantum critical (QC) regime as discussed
below. Here we have used the susceptibility due to the interacting
electrons themselves $4 \pi \chi = \varepsilon_{1} -
\varepsilon_{Si}$, namely the full measured dielectric constant,
$\varepsilon_{1}$, minus the static background dielectric constant
of the host silicon, $\varepsilon_{Si}$, to determine the
magnitude of the imaginary component of the conductivity shown in
Figure 3. The magnitude of the ratio becomes even larger and the
discrepancy greater if the full dielectric constant
$\varepsilon_{1}$ is considered instead.

The middle panel in Figure 3 shows the power $\alpha$ as
determined by Eq. (\ref{eq:alpha}). The values for Si:P are
approximately equal to, but slightly less than one, consistent
with Fig. 2. This indicates that the prefactor of the real and
imaginary components of the complex conductivity have the same
concentration dependence. When approaching the MIT, the frequency
dependence is expected to cross over to the QC
behavior\cite{Carini98,Henderson}, i.e. $\sigma_{1} \propto
\omega^{1/2}$ in NbSi, when the localization length $\xi$ becomes
comparable to the characteristic frequency dependent length scale,
e.g. the dephasing length, $\ell_{\omega}$\cite{Sondhi}. The
crossover is not a phase transition and need not be sharp,
therefore looking at a fixed window of frequencies, a broad,
smooth crossover from $\omega \rightarrow \omega^{1/2}$ would show
an averaged power of the frequency dependence similar to that
measured for NbSi shown in the middle panel of Figure 3. The fact
that we see an $\alpha \approx 1$ across our whole doping range in
Si:P, but an $\alpha$ that approaches 0.5 in NbSi indicates that
the critical regime in Si:P is much narrower. Simple dimensional
arguments\cite{Kivelson} give a result similar to the
non-interacting case\cite{Shapiro81} that the crossover should be
inversely proportional to the dopant DOS. The vastly smaller (a
factor of $10^{3}$) dopant density in Si:P relative to NbSi is
consistent with a narrower QC regime in Si:P as compared to NbSi.

\begin{figure}[tbh]
\centerline{\epsfig{figure=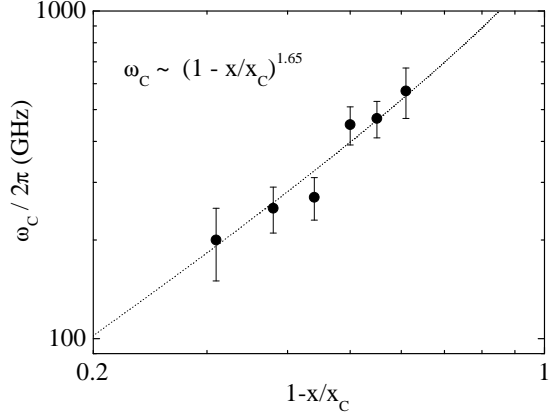,width=8cm}} \vspace{.2cm} \caption{This work's experimentally determined
crossover energy scale is represented by the full circles with a fit to a function of the relative concentration
$(1-x/x_{c})$ commonly used to normalize data from different materials.}
\end{figure}

Because our data spans a large range of concentrations, the doping
dependence of the crossover energy scale from linear to quadratic
can be analyzed to see whether its dependence is consistent with
the functional form of other energy scales, e.g. the Coulomb
interaction energy $U$ or the Coulomb gap width $\Delta$. Fig. 4
shows data of our experimentally determined crossover energy scale
plotted versus the commonly used functional form, $1-x/x_{c}$. M.
Lee et al. performed AC conductivity measurements on a pair of
Si:B samples \cite{MLee01}, and in said measurements, a sharp
crossover analogous to our own observations was seen. Our data
extends the observed range of the sharp crossover deep into the
insulating regime as M. Lee et al.'s original measurements were
closer to the critical concentration.

Recall that the Coulomb interaction energy between two sites
forming a resonant pair is $U(r_{\omega}) = e^{2}/\varepsilon_{1}
r_{\omega}$, which is dependent on concentration via the
dielectric constant and the localization length dependent most
probable hop distance. The Coulomb gap width as determined by ES
\cite{ES85} is $\Delta = e^{3} g_{0}^{1/2} /
\varepsilon_{1}^{3/2}$. Although M. Lee et al. postulated that a
smaller Coulomb gap governed the crossover, it is not unreasonable
to expect that this gap scales with the ES single particle one as
both are presumably caused by the long range Coulomb interaction.
The magnitude of the gap width is dependent on concentration
through the dielectric constant and through the DOS term. Thus it
is clear that a separate measurement of the full dielectric
constant, $\varepsilon_{1}$ as a function of concentration, as we
have done, need be performed in order to analyze the crossover
energy scale. Although there is no theoretical expectation for
$\omega_{c}$ or $\Delta$ to scale as a power law over the whole
doping range, we may still parameterize these quantities over our
doping range as power laws. We find that the crossover frequency
$\omega_{c}$ is proportional to $(1 - x/x_{c})^{1.65}$, and the
full dielectric constant $\varepsilon_{1} \propto (1 -
x/x_{c})^{-0.4}$. Using these values, it seems highly unlikely
that the $\hbar \omega_{c} = \Delta$, as the required DOS's
concentration dependence for this to be valid seems improbable.

We do find strong support that the crossover energy is determined
by the Coulomb interaction energy though. By setting the measured
crossover energy scale equal to the Coulomb interaction energy we
are able to determine both the magnitude of the localization
length and the exponent with which it diverges as a function of
concentration. Using an appropriate pre-factor for the overlap
integral \cite{Shklovskii}, $I_{0} = 10^{13} s^{-1}$ in the most
probable hop distance term, $r_{\omega}$, we find a localization
length dependence as plotted in the bottom panel of Fig. 3. We
find that the localization length exponent is close to unity, the
value originally predicted by McMillan in his scaling theory of
the MIT \cite{McMillan}, and the magnitude of the localization
length is reasonable. Both these results strongly point towards
the Coulomb interaction energy as being the energy scale at which
the observed frequency dependent crossover from ES to Mott-like
hopping conduction occurs.

In summary, we have observed behavior consistent with a Coulomb
glass across our entire range of doping concentrations in Si:P for
the frequency and concentration dependence of the real and
imaginary components of the conductivity. It is consistent with
theoretical predictions except in predicting the relative
magnitudes of $\sigma_{1}$ and $\sigma_{2}$ and the sharper than
expected crossover from linear to quadratic frequency dependence.
From the expected form of $U(x)$ and the reasonable behavior of
the extracted localization length, it seems likely that the
crossover is governed by the Coulomb interaction strength of a
resonant pair and not the Coulomb gap.

We would also like to thank Steve Kivelson as well as Boris
Shklovskii for useful conversations. This research was supported
by the National Science Foundation grant DMR-0102405.

\end{document}